\documentclass[12pt]{article}
\begin{document}
\title{Dark Matter, Gravity Waves, Higgs Bosons and other Elusive Entities}
\author{B.G. Sidharth \\
International Institute for Applicable Mathematics \& Information Sciences\\
Hyderabad (India) \& Udine (Italy)\\
B.M. Birla Science Centre, Adarsh Nagar, Hyderabad - 500 063
(India)}
\date{}
\maketitle
\begin{abstract}
The twentieth century has thrown up exotic concepts -- dark matter,
gravity waves, Higgs Bosons, Magnetic Monopoles and so on. The sad
truth is that even after several decades, these remain elusive to
observation and experiment. Some are now questioning these
conjectures. Their verification has become a matter more of hope
than conviction. We will examine some alternatives in the light of
the above pointing out that, on the other hand, an extra neutrino,
recently predicted by the author may have just been discovered.
\end{abstract}
\section{Dark Matter}
It is well known that F. Zwicky introduced the concept of dark
matter more than seventy five years ago to account for the anomalous
rotation curves of the galaxies \cite{narlikarcos,tduniv}. The
problem was that according to the usual Newtonian Dynamics the
velocities of the stars at the edges of galaxies should fall with
distance as in Keplarian orbits, roughly according to
\begin{equation}
v \approx \sqrt{\frac{GM}{r}}\label{e1}
\end{equation}
where $M$ is the mass of the galaxy, $r$ the distance from the
centre of the galaxy of the outlying star and $v$ the tangential
velocity of the star. Observations however indicated that the
velocity curves flatten out, rather than follow the law (\ref{e1}).
This necessitated the introduction of the concept of dark matter
which would take care of the discrepancy without modifying Newtonian
dynamics. However even after nearly eight decades, dark matter has
not been detected, even though there have been any number of
candidates proposed for this, for example SUSY particles, massive
neutrinos, undetectable brown dwarf stars,
even black holes and so on.\\
Very recent developments are even more startling. These concern the
rotating dwarf galaxies, which are satellites of the Milky Way
\cite{metz1,metz2}. These studies throw up a big puzzle. On the one
hand these dwarf satellites cannot contain any dark matter and on
the other hand the stars in the satellite galaxies are observed to
be moving much faster than predicted by Newtonian dynamics, exactly
as in the case of the galaxies themselves. Metz, Kroupa, Theis,
Hensler and Jerjen conclude that the only explanation lies in
rejecting dark matter and Newtonian gravitation. Indeed a well known
Astrophysicist, R. Sanders from the University of Groningen
commenting on these studies notes \cite{physorg}, "The authors of
this paper make a strong argument. Their result is entirely
consistent with the expectations of modified Newtonian dynamics
(MOND), but completely opposite to the predictions of the dark
matter hypothesis. Rarely is an observational test so definite."
Even more recently, dark matter has been ruled out by the
observational studies of the Kavli Institute in California, of the
interaction of electrons at the galactic edge with star light
\cite{petrosian}. Finally,
even more recent studies conclude that dark matter content has been vastly over estimated \cite{ag}.
Sawangwit and Shanks note that, apart from the fact that candidates for dark matter like weakly
interacting massive particles have not been detected so far, the dark matter problem in the Coma
cluster is a factor of a hundred less than when Zwicky first proposed it. This is due to the discovery of hot gas
in rich galaxy clusters. This makes the dark matter content only a factor of between four and five as a
discrepancy, rather than the original six hundred times.\\
In this note we point out that this could indeed be so, though not
via Milgrom's ad hoc modified dynamics \cite{mil1,mil2,tduniv},
according to which a test particle at a distance $r$ from a large
mass $M$ is subject to the acceleration $a$ given by
\begin{equation}
a^2/a_0 = MGr^{-2},\label{3em1}
\end{equation}
where $a_0$ is an acceleration such that standard Newtonian dynamics
is a good approximation only for accelerations much larger than
$a_0$. The above equation however would be true when $a$ is much
less than $a_0$. Both the statements in (\ref{3em1}) can be combined
in the heuristic relation
\begin{equation}
\mu (a/a_0) a = MGr^{-2}\label{3em2}
\end{equation}
In (\ref{3em2}) $\mu(x) \approx 1$ when $x >> 1, \, \mbox{and}\,
\mu(x) \approx x$ when $x << 1$. It must be stressed that
(\ref{3em1}) or (\ref{3em2}) are not deduced from any theory, but
rather are an ad hoc prescription to explain observations.
Interestingly it must be mentioned that most of the implications of
Modified Newtonian Dynamics or MOND do not
depend strongly on the exact form of $\mu$.\\
It can then be shown that the problem of galactic velocities is now
solved \cite{mil1,mil2,mil3,mil4,mil5}. Nevertheless, most
physicists are not comfortable with MOND because of the ad hoc
nature of (\ref{3em1}) and (\ref{3em2}).\\
We now come to the cosmological model described by the author in
1997 (Cf.ref.\cite{ijmpa1998,tduniv} and several references
therein), in which the universe, under the influence of dark energy
would be accelerating with a small acceleration. Several other
astrophysical relations, some of them hitherto inexplicable such as
the Weinberg formula giving the pion mass in terms of the Hubble
constant were also deduced in this model (Cf.also ref.\cite{cu} and
references therein). While all this was exactly opposite to the then
established theory, it is well known that the picture was
observationally confirmed soon thereafter through the work of
Perlmutter and others (Cf.ref.\cite{cu}). Interestingly, in this
model Newton's
gravitational constant varied inversely with time.\\
Cosmologies with time varying $G$ have been considered in the past,
for example in the Brans-Dicke theory or in the Dirac large number
theory or by Hoyle \cite{barrowparsons,narfpl,narburbridge,5,6}. In
the case of the Dirac cosmology, the motivation was Dirac's
observation that the supposedly large number coincidences involving
$N \sim 10^{80}$, the number of elementary particles in the universe
had an underlying message if it is recognized that
\begin{equation}
\sqrt{N} \propto T\label{3ea1}
\end{equation}
where $T$ is the age of the universe. Equation (\ref{3ea1}) too
leads to a $G$ decreasing inversely
with time as we will now show. We follow a route slightly different from that of Dirac.\\
From (\ref{3ea1}) it can easily be seen that
\begin{equation}
T = \sqrt{N} \tau\label{5}
\end{equation}
where $\tau$ is a typical Compton time of an elementary particle
$\sim 10^{-23}secs$, because $T$, the present age of the universe is
$\sim 10^{17}secs$. We also use the following well known relation
which has been obtained a long time ago through different routes
\cite{ruffinizang,hayakawa,nottalefractal}
\begin{equation}
R = \frac{2GM}{c^2}\label{8}
\end{equation}
Further multiplying both sides of (\ref{5}) by $c$ we get the famous
Weyl-Eddington relation
\begin{equation}
R = \sqrt{N} l\label{9}
\end{equation}
where $l = c\tau$ is a typical Compton length $\sim 10^{-13}cms$. We
will also use another well known relation (Cf. above references),
$$M = Nm,$$
where $m$ is a typical elementary particle mass, like that of the
pion, $10^{-25}gm$.\\
Use of (\ref{9}) and the above in (\ref{8}) now leads to
\begin{equation}
G = \frac{c^2 l}{2\sqrt{N}m} = \left(\frac{c^2 l \tau}{2m}\right)
\cdot \frac{1}{T} \equiv \frac{G_0}{T}\label{10}
\end{equation}
Equation (\ref{10}) gives the above stated inverse dependence of the
gravitational constant $G$ on time, which Dirac obtained. On the
other hand this same relation was obtained by a different route in
the author's dark energy -- fluctuations cosmology in 1997. This
work, particularly in the context of the Planck scale has been there
for many years in the literature (Cf.\cite{cu,tduniv,uof} and
references therein). Suffice to say that all the supposedly so
called accidental Large Number Relations like (\ref{9}) as also the
inexplicable Weinberg formula which relates the Hubble constant to
the mass of a pion, follow as deductions in this
cosmology. The above references give a comprehensive picture.\\
In any case, our starting point is, equation (\ref{10}) where $T$ is
time (the age of the universe) and $G_0$ is a constant. Furthermore,
other routine effects like the precession of the perihelion of
Mercury and the bending of light and so on are also explained with
(\ref{10}) as will be briefly discussed below.  We will also see
that there is observational evidence for (\ref{10}) (Cf. also
\cite{uzan} which described various observational evidences for the
variation of $G$, for example from solar system observations, from
cosmological observations and even from the
palaeontological studies point of view).\\
With this background, we now mention some further tests for equation
(\ref{10}).\\
This could explain the other General Relativistic effects like the
shortening of the period of binary pulsars and so on
(Cf.ref.\cite{cu,tduniv,bgsnc115b,bgsfpl} and other references
therein). Moreover, we could now also explain, the otherwise
inexplicable anomalous acceleration of the Pioneer space crafts
(Cf.ref.\cite{tduniv} for details). We will briefly revisit some of these effects later.\\
We now come to the problem of galactic rotational curves mentioned
earlier (cf.ref.\cite{narlikarcos}). We would expect, on the basis
of straightforward dynamics that the rotational velocities at the
edges of galaxies would fall off according to
\begin{equation}
v^2 \approx \frac{GM}{r}\label{3ey33}
\end{equation}
which is (\ref{e1}). However it is found that the velocities tend to
a constant value,
\begin{equation}
v \sim 300km/sec\label{3ey34}
\end{equation}
This, as noted, has lead to the postulation of the as yet undetected
additional matter alluded to, the so called dark matter.(However for
an alternative view point Cf.\cite{sivaramfpl93}). We observe that
(\ref{10}) can be written for an increase $t$,in time, small
compared to the age of the universe, now written as $t_0$
\begin{equation}
G = \frac{G_0}{t_0 + t} = \frac{G_0}{t_0}  \left(1 -
\frac{t}{t_0}\right)\label{3.17}
\end{equation}
Using (\ref{3.17}), let us consider the gravitational potential
energy $V$ between two masses, $m_1$ and $m_2$ by:
\begin{equation}
V = \frac{G m_1 m_2}{r_0} = \frac{G_0}{t_0} \cdot
\frac{m_1m_2}{r_0}\label{14}
\end{equation}
After a time $t$ this would be, by (\ref{3.17}),
\begin{equation}
V = \frac{G_0}{t_0} \left(1 - \frac{t}{t_0}\right)
\frac{m_1m_2}{r}\label{15}
\end{equation}
Equating (\ref{14}) and (\ref{15}) we get,
\begin{equation}
r = r_0 \left(\frac{t_0}{t_0 + t}\right) \left(= r_0
(1-\frac{t}{t_0})\right)\label{3.18}
\end{equation}
The relation (\ref{3.18}) has been deduced by a different route by
Narlikar \cite{narlikarcos}.  From (\ref{3.18}) it easily follows
that,
\begin{equation}
a \equiv (\ddot{r}_{o} - \ddot{r}) \approx \frac{1}{t_o}
(t\ddot{r_o} + 2\dot r_o) \approx -2 \frac{r_o}{t^2_o}\label{3ey35}
\end{equation}
as we are considering intervals $t$ small compared to the age of the
universe and nearly circular orbits. In (\ref{3ey35}), $a$ or the
left side of (\ref{3ey35}) gives the new extra effect due to
(\ref{3.17}) and (\ref{3.18}), this being a departure from the usual
Newtonian gravitation. Equation (\ref{3ey35}) shows
(Cf.ref\cite{bgsnc115b} also) that there is an anomalous inward
acceleration, as if there is
an extra attractive force, or an additional central mass.\\
So, introducing the extra acceleration (\ref{3ey35}), we get,
\begin{equation}
\frac{GMm}{r^2} + \frac{2mr}{t^2_o} \approx
\frac{mv^2}{r}\label{3ey36}
\end{equation}
From (\ref{3ey36}) it follows that
\begin{equation}
v \approx \left(\frac{2r^2}{t^2_o} + \frac{GM}{r}\right)^{1/2}
\label{3ey37}
\end{equation}
So (\ref{3ey37}) replaces (\ref{e1}) in this model. This shows that
as long as
\begin{equation}
\frac{2r^2}{t_0^2} < < \frac{GM}{r},\label{20}
\end{equation}
Newtonian dynamics holds. But when the first term on the left side
of (\ref{20}) becomes of the order of the second (or greater), the
new dynamical effects come in.\\
For example from (\ref{3ey37}) it is easily seen that at distances
well within the edge of a typical galaxy, that is $r < 10^{23}cms$
the usual equation (\ref{3ey33}) holds but as we reach the edge and
beyond, that is for $r \geq 10^{24}cms$ we have $v \sim 10^7 cms$
per second, in agreement with (\ref{3ey34}). In fact as can be seen
from (\ref{3ey37}), the first term in the square root has an extra
contribution (due to the varying $G$) which exceeds the second term
as we approach the galactic edge, as if there is an extra mass, that
much more.\\
We would like to stress that the same conclusions will apply to the
latest observations of the satellite galaxies (without requiring any
dark matter).\\
Let us for example consider the Megallanic clouds \cite{stave}. In
this case, as we approach their edges, the first term within the
square root on the right side of (\ref{3ey37}) or the left term of
(\ref{20}) already becomes of the order of the
second term, leading to the new non Newtonian effects.\\
The point is that the above varying $G$ scheme described in
(\ref{10}) or (\ref{3ey37}) reproduces all the effects of General
Relativity as noted above, as also the anomalous acceleration of the
Pioneer space crafts in addition to the conclusions of MOND
regarding an alternative for dark matter, and is applicable to the
latest observations of satellite galaxies. The satellite galaxy
rotation puzzle is thus resolved.
\section{General Relativity}
It is now accepted that the tests which General Relativity proposed
have all been observed, except for gravitational waves which have
eluded detection for nearly a century. We would like to observe that
the variation of the given constant mentioned earlier can also
explain the various tests of General Relativity. Let us see how. We
first observe that our starting point is the precession of the
perihelion of mercury. We observe that from (\ref{10}) or
(\ref{3.17}) it follows that \cite{narlikarcos}
\begin{equation}
G = G_o (1- \frac{t}{t_o})\label{3ey15}
\end{equation}
where $G_o$ is the present value of $G$ and $t_o$ is the present age
of the Universe and $t$ the (relatively small) time elapsed from the
present epoch. Similarly one could deduce that
(cf.ref.\cite{narlikarcos}),
\begin{equation}
r = r_o \left(\frac{t_o}{t_o+t}\right)\label{3ey16}
\end{equation}
We next use Kepler's Third law\cite{gold}:
\begin{equation}
\tau = \frac{2 \pi a^{3/2}}{\sqrt{GM}}\label{3ey17}
\end{equation}
$\tau$ is the period of revolution, $a$ is the orbit's semi major
axis, and $M$ is the mass of the sun. Denoting the average angular
velocity of the planet by
$$\dot \Theta \equiv \frac{2 \pi}{\tau},$$
it follows from (\ref{3ey15}), (\ref{3ey16}) and (\ref{3ey17}) that
$$\dot \Theta - \dot \Theta_o =  \dot \Theta_0 \frac{t}{t_o},$$
where the subscript $o$ refers to the present epoch.\\
Whence,
\begin{equation}
\omega (t) \equiv \Theta - \Theta_o =  \frac{\pi}{\tau_o t_o}
t^2\label{3ey18}
\end{equation}
Equation (\ref{3ey18}) gives the average perihelion precession at
time '$t$'. Specializing to the case of Mercury, where $\tau_o =
0.25$ year, it follows from (\ref{3ey18}) that the average
precession per year at time '$t$' is given by
\begin{equation}
\omega (t) =  \frac{4\pi t^2}{t_0}\label{3ey19}
\end{equation}
Whence, considering $\omega (t)$ for years $t=1,2, \cdots , 100,$ we
can obtain from (\ref{3ey19}), the correct perihelion precession per
century as \cite{bgsnc115b},
$$\omega = \sum^{100}_{n=1} \omega (n) \approx 43'' ,$$
if the age of the universe is taken to be $\approx 2 \times 10^{10}$ years.\\
Conversely, if we use the observed value of the precession in
(\ref{3ey19}),
we can get back the above age of the universe.\\
Interestingly it can be seen from (\ref{3ey19}), that the precession depends on the epoch.\\
We can similarly demonstrate that orbiting objects will have an
anamolous inward
radial acceleration.\\
For $r \sim 10^{14}cm$, as is the case of the space crafts Pioneer
$10$ or Pioneer $11$, this gives,
$a_r \geq 10^{-11}cm/sec^2$ This can be further refined to $a_\gamma \leq 10^{-10}cm$.\\
Interestingly Anderson et al.,\cite{andersongrqc} claim to have
observed an anomalous inward acceleration
of $\sim 10^{-8} cm/sec^2$ for more than a decade.\\
We could also explain the correct gravitational bending of light in
the same vein.\\
The inexplicable anomalous accelerations of the Pioneer spacecrafts
already alluded to, which have been observed by J.D. Anderson and
coworkers at the Jet Propulsion Laboratory for well over a decade
\cite{andersonphysrev,andersongrqc} have posed a puzzle. This can be
explained, in a simple way as follows: In fact from the usual
orbital equations we have \cite{bgstest}
$$v \dot {v} \approx -\frac{GM}{2tr} (1 + e cos \Theta )-\frac{GM}{r^2} \dot {r}(1+e cos \Theta )$$
$v$ being the velocity of the spacecraft and $t$ is the time in
general. It must be observed that the first term on the right side
is the new effect due to (\ref{3ea2}). There is now an anomalous
acceleration given by
$$a_r = \langle \dot v \rangle_{\mbox{anom}} = \frac{-GM}{2t r v} (1+e cos \Theta )$$
$$\approx -\frac{GM}{2t\lambda} (1+e)^3$$
where
$$\lambda = r^4 \dot \Theta^2$$
If we insert the values for the Pioneer spacecrafts we get
$$a_r \sim -10^{-7} cm/sec^2$$
This is the anomalous acceleration reported by Anderson and co-workers.\\
We will next deduce that this case also explains correctly the
observed decrease in the orbital period of the binary pulsar $PSR\,
1913 + 16$, which has also been attributed to as yet
undetected gravitational waves \cite{davis}.\\
It should also  be remarked that in the case of gravitational
radiation, there are some objections relevant to the
calculation (Cf.ref.\cite{davis}).\\
Finally, we may point out that a similar shrinking in size with time
can be expected of galaxies themselves, and in general,
gravitationally bound systems. We will see a special case for the
solar system.\\
To consider the above result in a more general context, we come back
to the well known orbital equation \cite{bgstest}
\begin{equation}
d^2 u/d\Theta^2 + u = \mu_0/h^2\label{3ea14}
\end{equation}
where $\mu_0 = GM$ and $u$ is the usual inverse of radial distance.\\
$M$ is the mass of the central object and $h = r^2 d\Theta / dt$ - a
constant. The solution of (\ref{3ea14}) is well known,
$$lu = 1 + ecos \Theta$$
where $l = h^2/\mu_0$.\\
It must be mentioned that in the above purely classical analysis,
there is no
precession of the perihelion.\\
\indent We now replace $\mu_0$ by $\mu$ and also assume $\mu$ to be
varying slowly because $G$ itself varies slowly and uniformly, as
noted earlier:
\begin{equation}
\dot{\mu} = d\mu / dt = K, \mbox{a \, constant}\label{3ea15}
\end{equation}
remembering that $\dot{K} \sim 0 (1/T^2)$ and so can be neglected.\\
\indent Using (\ref{3ea15}) in (\ref{3ea14}) and solving the orbital
equation (\ref{3ea14}), the solution can now be obtained as
\begin{equation}
u = 1/l + (e/l)cos\Theta + K l^2 \Theta/h^3 + Kl^2 e\Theta cos\Theta
/h^3\label{3ea16}
\end{equation}
Keeping terms up to the power of '$e$' and $(K/\mu_0 )^2$, the time
period '$\tau$' for one revolution is given to this order of
approximation by
\begin{equation}
\tau = 2\pi L^2/h\label{3ea17}
\end{equation}
From (\ref{3ea16})
\begin{equation}
L  = l - \frac{Kl^4\Theta}{h^3}\label{3ea18}
\end{equation}
Substituting (\ref{3ea18}) in (\ref{3ea17}) we have
\begin{equation}
\tau = \frac{2\pi}{h} \left(l^2 -
\frac{2Kl^5\Theta}{h^3}\right)\label{3ea19}
\end{equation}
The second term in (\ref{3ea19}) represents the change in time
period for one revolution. The decrease of time period is given by
\begin{equation}
\delta \tau = 8\pi^2 l^3 K/\mu_0^2\label{3ea20}
\end{equation}
The second term in (\ref{3ea18}) indicates the decrease in latus-rectum.\\
\indent For one revolution the change of latus-rectum is given by
\begin{equation}
\delta l = 2\pi Kl^{2.5}/\mu^{1.5}_0\label{3ea21}
\end{equation}
In the solar system, we have,
$$K = 898800\, cm \, gm$$
Using $K$ and $\mu_0$ to find the change in time period and the
latus rectum in the varying $G$ case by substituting in
(\ref{3ea20}) and (\ref{3ea21}) respectively for Mercury we get
$$\delta T = 1.37 \times 10^{-5} sec/rev$$
\begin{equation}
\delta l = 4.54 cm/rev\label{3ea22}
\end{equation}
We observe that the equations (\ref{3ea20}), (\ref{3ea21}) or
(\ref{3ea22}) show a decrease in distance and in the time of
revolution. If we use for the planetary motion, the General
Relativistic analogue of (\ref{3ea14}), viz.,
$$\frac{d^2 u}{d\Theta^2} + u = \frac{\mu_0}{h^2} (1 + 3h^2 u^2),$$
then while we recover the precession of the perihelion of Mercury,
for example, there is no effect similar to (\ref{3ea20}),
(\ref{3ea21}) or (\ref{3ea22}). On the other hand this effect is
very minute-- just a few centimeters per year in the case of the
earth-- and only protracted careful observations can detect it.
Moreover these changes could also
be masked at least partly, by gravitational and other perturbations.\\
However as noted, the decrease of the period in (\ref{3ea20}) has
been observed in the case of Binary Pulsars.\\
Finally, it has been pointed out that (\ref{3ey15}) itself can be
shown to yield an effect that simulates gravitational waves without
invoking the full General Relativity \cite{tduniv}.
\section{Higgs Bosons}
We now come to another aspect. According to the standard model of
particle physics we need the Higgs Bosons to provide mass to the
various particles in the universe. However these Bosons, first
suggested by Peter Higgs in 1965 have eluded detection for decades.
More recently, hopes that these so called God particles would be
detected at the LHC have received a segment, because, the Tevatron
rules out the existence of Higgs particles in the expected range
below $200 GeV$. Physicists however are still hoping that the LHC
would throw up the Higgs, though some are now veering to the view
that physics beyond the
standard model would be more interesting without the Higgs.\\
In this latter context, we will now argue that it is possible for
both \index{electromagnetism}electromagnetism and
\index{gravitation}gravitation to emerge from a gauge like
formulation \cite{bgsgrav2}. In Gauge Theory, which as we saw in
Chapter 2, is a Quantum Mechanical generalization of
\index{Weyl}Weyl's original geometry, we generalize, as is well
known, the original phase transformations, which are global with the
phase $\lambda$ being a constant, to local phase transformations
with $\lambda$ being a function of the coordinates \cite{jacob}. As
is well known this leads to a covariant gauge derivative. For
example, the transformation arising from $(x^\mu) \to (x^\mu +
dx^\mu)$,
\begin{equation}
\psi \to \psi e^{-\imath \lambda}\label{Ee1}
\end{equation}
leads to the familiar \index{electromagnetic}electromagnetic
potential gauge,
\begin{equation}
A_\mu \to A_\mu - \partial_\mu \lambda\label{Ee2}
\end{equation}
The above transformation, ofcourse, is a \index{symmetry}symmetry transformation.
In the transition from (\ref{Ee1}) to (\ref{Ee2}), we expand the exponential, retaining terms only to the first
order in coordinate differentials.\\
Let us now consider the case where there is a minimum cut off in the
spactime intervals. As we saw this leads to a
\index{noncommutative}noncommutative geometry
(Cf.ref.\cite{annales})
\begin{equation}
[dx_\mu , dx_\nu ] = O(l^2)\label{Ee3}
\end{equation}
where $l$ is the minimum scale. From (\ref{Ee3}) it can be seen that
if $O(l^2)$ is neglected, we are back with the familiar commutative
\index{spacetime}spacetime. The new effects of \index{fuzzy
spacetime}fuzzy spacetime arise when the right side of (\ref{Ee3})
is not neglected. Based on this we had argued in Chapter 5 that it
is possible to reconcile \index{electromagnetism}electromagnetism
and \index{gravitation}gravitation \cite{bgsgrav,ax,fpl1,br2}. If in
the transition from (\ref{Ee1} to (\ref{Ee2}) we retain, in view of
(\ref{Ee3}), squares of differentials, in the
\index{expansion}expansion of the function $\lambda$ we will get
terms like
\begin{equation}
\left\{ \partial_\mu \lambda \right\} dx^\mu + \left(\partial_\mu
\partial_\nu + \partial_\nu \partial_\mu \right) \lambda \cdot
dx^\mu dx^\nu\label{Ee4}
\end{equation}
where we should remember that in view of (\ref{Ee3}), the derivatives
(or the product of coordinate differentials) do not commute as indeed we saw in Chapter 5.
As in the usual theory the coefficient of $dx^\mu$ in the first term of (\ref{Ee4}) represents now,
not the gauge term but the \index{electromagnetic}electromagnetic potential itself: Infact, in this
\index{noncommutative}noncommutative geometry, it can be shown that this \index{electromagnetic}electromagnetic
potential reduces to the potential in \index{Weyl}Weyl's original gauge theory \cite{bgsgrav2,bgsgrav}.\\
Without the noncommutativity, the potential $\partial_\mu \lambda$ would lead to a vanishing
\index{electromagnetic}electromagnetic field. However as we saw \index{Dirac}Dirac pointed out
in his famous \index{monopole}monopole paper in 1930 that a non integrable phase $\lambda (x,y,z)$ leads
as above directly to the \index{electromagnetic}electromagnetic potential, and moreover this was an alternative
formulation of the original \index{Weyl}Weyl theory \cite{dirac4,nc}.\\
Returning to (\ref{Ee4}) we identify the next coefficient with the
\index{metric tensor}metric tensor giving the
\index{gravitation}gravitational field:
\begin{equation}
ds^2 = g_{\mu \nu} dx^\mu dx^\nu = \left(\partial_\mu \partial_\nu +
\partial_\nu \partial_\mu \right) \lambda dx^\mu dx^\nu\label{Ee5}
\end{equation}
Infact one can easily verify that $ds^2$ of (\ref{Ee5}) is an
invariant. We now specialize to the case of the linear theory in
which squares and higher powers of the deviation from the
\index{Minkowski}Minkowski metric,  $h^{\alpha \beta}$ can be
neglected. In this case it can easily be shown that
\begin{equation}
2 \Gamma^\beta_{\mu \nu} = h_{\beta  \mu ,\nu} + h_{\nu \beta ,\mu}
- h_{\mu \nu ,\beta}\label{Ee6}
\end{equation}
where in (\ref{Ee6}), the $\Gamma$s denote Christofell symbols. From
(\ref{Ee6}) by a contraction we have
\begin{equation}
2\Gamma^\mu_{\mu \nu} = h_{\mu \nu ,\mu} = h_{\mu \mu ,
\nu}\label{Ee7}
\end{equation}
If we use the well known gauge condition \cite{ohanian}
$$\partial_\mu \left(h^{\mu \nu} - \frac{1}{2} \eta^{\mu \nu} h_{\mu \nu}\right) = 0, \, \mbox{where}\, h = h^\mu_\mu$$
then we get
\begin{equation}
\partial_\mu h_{\mu \nu} = \partial_\nu h^\mu_\mu \equiv \partial_\nu h\label{Ee8}
\end{equation}
(\ref{Ee8}) shows that we can take the $\lambda$ in (\ref{Ee4}) as $\lambda = è$, both for the
\index{electromagnetic}electromagnetic potential $A_\mu$ and the \index{metric tensor}metric
tensor $h_{\mu \nu}$. (\ref{Ee7}) further shows that the $A_\mu$ so defined becomes identical to
\index{Weyl}Weyl's gauge invariant potential \cite{berg}.\\
However it is worth reiterating that in the present formulation, we have a
\index{noncommutative}noncommutative geometry, that is the derivatives do not commute and moreover
we are working to the order where $l^2$ cannot be neglected. Given this condition both the
\index{electromagnetic}electromagnetic potential and the \index{gravitation}gravitational potential
are seen to follow from the gauge like theory. By retaining coordinate differential squares, we are
even able to accommodate apart from the usual \index{spin}spin 1 gauge particles, also the \index{spin}spin 2
graviton which otherwise cannot be accommodated in the usual gauge theory. If however $O(l^2) = 0$, then
we are back with commutative \index{spacetime}spacetime, that is a usual point \index{spacetime}spacetime
and the usual gauge theory describing \index{spin}spin 1 particles.\\
We had reached this conclusion in Chapter 5 (Cf. ref. \cite{annales}), though from a different, non gauge
point of view. The advantage of the present formulation is that it provides a transparent link with conventional
theory on the one hand, and shows how the other interactions described by non Abelian gauge theories smoothly fit into the picture.\\
Finally it may be pointed out that we had already argued that a
\index{fuzzy spacetime}fuzzy spacetime input explains why the purely
classical \index{Kerr-Newman}Kerr-Newman metric gives the purely
Quantum Mechanical anomalous \index{gyromagnetic ratio}gyromagnetic
ratio of the \index{electron}electron \cite{ar4,sakharov}, thus
providing a link between \index{General Relativity}General
Relativity and \index{electromagnetism}electromagnetism. This
provides further support to the above considerations.\\
Let us now return to the \index{gauge field}gauge field itself. As
is well known, this could be obtained as a generalization of the
above phase function $\lambda$ to include fields with internal
degrees of freedom. For example $\lambda$ could be replaced by
$A_\mu$ given by \cite{moriyasu}
\begin{equation}
A_\mu = \sum_{\imath} A^\imath_\mu (x)L_\imath ,\label{Eex1}
\end{equation}
The \index{gauge field}gauge field itself would be obtained by using
Stoke's Theorem and (\ref{Eex1}). This is a very well known
procedure: considering a circuit, which for simplicity we can take
to be a parallelogram of side $dx$ and $dy$ in two dimensions, we
can easily deduce the equation for the field, viz.,
\begin{equation}
F_{\mu \nu} = \partial_\mu A_\nu - \partial_\nu A_\mu - \imath q
[A_\mu , A_\nu ],\label{Eex2}
\end{equation}
$q$ being the \index{gauge field}gauge field coupling constant.\\
In (\ref{Eex2}), the second term on the right side is typical of a non Abelian \index{gauge field}gauge field.
In the case of the\index{U(1)}(U(1) \index{electromagnetic}electromagnetic field, this latter term vanishes.\\
Further as is well known, in a typical Lagrangian like
\begin{equation}
\mathit{L} = \imath \bar \psi \gamma^\mu D_\mu \psi - \frac{1}{4}
F^{\mu \nu} F_{\mu \nu} - m \bar \psi \psi\label{Eex3}
\end{equation}
$D$ denoting the Gauge \index{covariant derivative}covariant derivative, there is no \index{mass}mass term for the
field \index{Boson}Bosons. Such a \index{mass}mass term in (\ref{Eex3}) must have the form $m^2 A^\mu A_\mu$ which
unfortunately is not Gauge invariant.\\
This as we saw in Chapter 2, was the shortcoming of the original \index{Yang-Mills}Yang-Mills Gauge Theory: The Gauge
Bosons would be \index{mass}massless and hence the need for a \index{symmetry breaking}symmetry breaking, \index{mass}mass generating mechanism.\\
The well known remedy for the above situation has been to consider,
in analogy with \index{superconductivity}superconductivity theory,
an extra phase of a self coherent system (Cf.ref.\cite{moriyasu} for
a simple and elegant treatment and also refs. \cite{jacob} and
\cite{taylor}). Thus instead of the \index{gauge field}gauge field
$A_\mu$, we consider a new phase adjusted \index{gauge field}gauge
field after the \index{symmetry}symmetry is broken
\begin{equation}
W_\mu = A_\mu - \frac{1}{q} \partial_\mu \phi\label{Eex4}
\end{equation}
The field $W_\mu$ now generates the \index{mass}mass in a self
consistent manner via a Higgs mechanism. Infact the kinetic energy
term
\begin{equation}
\frac{1}{2} |D_\mu \phi |^2\quad ,\label{Eex5}
\end{equation}
where $D_\mu$ in (\ref{Eex5}) denotes the Gauge \index{covariant
derivative}, now becomes
\begin{equation}
|D_\mu \phi_0 |^2 = q^2|W_\mu |^2 |\phi_0 |^2 \, ,\label{Eex6}
\end{equation}
Equation (\ref{Eex6}) gives the \index{mass}mass in terms of the ground state $\phi_0$.\\
The whole point is as follows: The \index{symmetry breaking}symmetry breaking of the \index{gauge field}gauge
field manifests itself only at short length scales signifying the fact that the field is mediated by particles with large
\index{mass}mass. Further the internal \index{symmetry}symmetry space of the \index{gauge field}gauge field is broken by an
external constraint: the wave function has an intrinsic relative phase factor which is a different function of spacetime
coordinates compared to the phase change necessitated by the minimum coupling requirement for a free particle with the gauge potential.
This cannot be achieved for an ordinary point like particle, but a new type of a physical system, like the self coherent system of
\index{superconductivity}superconductivity theory now interacts with the \index{gauge field}gauge field. The second or extra term in
(\ref{Eex4}) is effectively an external field, though (\ref{Eex6}) manifests itself only in a relatively small spatial interval.
The $\phi$ of the Higgs field in (\ref{Eex4}), in analogy with the phase function of  \index{Cooper pairs}Cooper pairs of
\index{superconductivity}superconductivity theory comes with a \index{Landau-Ginzburg}Landau-Ginzburg potential $V(\phi)$.\\
Let us now consider in the \index{gauge field}gauge field
transformation, an additional phase term, $f(x)$, this being a
scalar. In the usual theory such a term can always be gauged away in
the \index{U(1)}U(1) \index{electromagnetic}electromagnetic group.
However we now consider the new situation of a
\index{noncommutative}noncommutative geometry referred to above,
\begin{equation}
\left[dx^\mu , dx^\nu \right] = \Theta^{\mu \nu} \beta , \beta \sim
0 (l^2)\label{Eex7}
\end{equation}
where $l$ denotes the minimum \index{spacetime}spacetime cut off.
Equation (\ref{Eex7}) is infact \index{Lorentz}Lorentz covariant.
Then the $f$ phase factor gives a contribution to the second order
in coordinate differentials,
$$\frac{1}{2} \left[\partial_\mu B_\nu - \partial_\nu B_\mu \right] \left[dx^\mu , dx^\nu \right]$$
\begin{equation}
+ \frac{1}{2} \left[\partial_\mu B_\nu + \partial_\nu B_\mu \right]
\left[dx^\mu dx^\nu + dx^\nu dx^\mu \right]\label{Eex8}
\end{equation}
where $B_\mu \equiv \partial_\mu f$.\\
As can be seen from (\ref{Eex8}) and (\ref{Eex7}), the new
contribution is in the term which contains the commutator of the
coordinate differentials, and not in the symmetric second term.
Effectively, remembering that $B_\mu$ arises from the scalar phase
factor, and not from the non-Abelian \index{gauge field}gauge field,
in equation (\ref{Eex2}) $A_\mu$ is replaced by
\begin{equation}
A_\mu \to A_\mu + B_\mu = A_\mu + \partial_\mu f\label{Eex9}
\end{equation}
Comparing (\ref{Eex9}) with (\ref{Eex4}) we can immediately see that the effect of noncommutativity is
precisely that of providing a new \index{symmetry breaking}symmetry breaking term to the \index{gauge field}gauge field,
instead of the $\phi$ term, (Cf.refs. \cite{cr39,ijmpe}) a term not belonging to the \index{gauge field}gauge field itself.\\
On the other hand if we neglect in (\ref{Eex7}) terms $\sim l^2$,
then there is no extra contribution coming from (\ref{Eex8}) or
(\ref{Eex9}), so that we are in the usual non-Abelian \index{gauge
field}gauge field theory, requiring a broken
\index{symmetry}symmetry to obtain an equation like (\ref{Eex9}).
This is not surprising because as noted several times if we neglect
the term $\sim l^2$ in (\ref{Eex7}) then we are back with the usual
commutative theory and the usual \index{Quantum Mechanics}Quantum
Mechanics.\\
The matters have been dealt with more recently too \cite{bgsijmpe}.
\section{Miscellaneous Remarks}
The above consideration of non-commutative spacetime have also been
shown to lead to the conclusion that the magnetic monopole is
redundant, as indeed Dirac himself had suggested \cite{nc,tduniv}.
On the other hand, based on recent work with ultra high energy
fermions, the author had suggested recently \cite{arxiv} that there
would be an extra neutrino (rather like a sterile neutrino). It is
quite remarkable that researches at Fermi Lab have just confirmed
that indeed such a fourth flavour neutrino exists \cite{roe}.\\
Further, Einstein's General Relativity deals with gravitation and
its unification with Quantum Mechanics or electromagnetism is still
eluding us, even after a century. In this connection, We would first
like to briefly touch upon the author's Planck oscillator model
which, over the years, successfully describe phenomena from an
elementary
particle to the universe itself.\\
We can easily verify that the Planck scale $l$ plays the role of the
Compton length and the Schwarzchild radius of a black hole of the
mass $m$ \cite{kiefer}
\begin{equation}
l = \frac{\hbar}{2mc}, \, l = \frac{2Gm}{c^2} m \sim 10^{-5}gm \, l
\sim 10^{-33}cm\label{ee2}
\end{equation}
Today in various Quantum gravity approaches the Planck length $l$ is
considered to be the fundamental minimum length, and so also the
time interval $t = l/c$. Spacetime intervals smaller than given in
(\ref{ee2}) are meaningless both classically and Quantum
mechanically. Classically because we cannot penetrate the
Schwarzchild radius, and Quantum mechanically because we encounter
unphysical phenomena inside a typical Compton scale. We will return
to this point but all this has been discussed in greater detail by
the author and others (Cf.ref.\cite{tduniv} and several references
therein).\\
At another level, it may be mentioned that the author's 1997
cosmological model invoked a background dark energy and fluctuations
therein to deduce a model of the universe that was accelerating with
a small cosmological constant, together with several other relations
completely consistent with Astrophysics and Cosmology
(Cf.ref.\cite{cu} and several references therein). The observations
of distant supernovae by Perlmutter and others confirmed in 1998 the
dark energy driven accelerating universe of
the author. All this is well known.\\
It is against this backdrop that the author had put forward his
model of Planck oscillators in the dark energy driven Quantum
vacuum, several years ago (Cf.ref.\cite{uof} and several references
therein, \cite{fpl2000}). Let us consider an array of $N$ particles,
spaced a distance $\Delta x$ apart, which behave like oscillators
that are connected by springs. As is known we then have
\cite{uof,fpl152002}
\begin{equation}
r = \sqrt{N \Delta x^2} ka^2 \equiv k \Delta x^2 = \frac{1}{2} k_B
T\label{ee3}
\end{equation}
where $k_B$ is the Boltzmann constant, $T$ the temperature, $r$ the
total extension and $k$ is the spring constant given by
\begin{equation}
\omega_0^2 = \frac{k}{m}\label{ee4}
\end{equation}
\begin{equation}
\omega = \left(\frac{k}{m} a^2\right)^{\frac{1}{2}} \frac{1}{2} =
\omega_0 \frac{a}{r}\label{ee5}
\end{equation}
In (\ref{ee4}) $\omega_0$ is the frequency of the individual
oscillator, while in (\ref{ee5}) $\omega$ is the frequency of the
array of $N$ oscillators, $N$
given in (\ref{ee3}).\\
We can easily show from the above theory of oscillators that an
oscillator with the Planck mass and with a spatial extent at the
Planck scale has the same temperature as the Beckenstein temperature
of a Schwarzchild Black Hole of mass given by the Planck mass. The
above results can be obtained by a different route as described in
\cite{bgsijmpa}.\\
It has also been shown that, given the well known effect that the
universe consists of $N \sim 10^{80}$ elementary particles like the
pion, it is possible to deduce that a typical elementary particle
consists of $n \sim 10^{40}$ Planck oscillators. These form a
coherent array unlike in string theory, where we deal with single
oscillators. Briefly, to recapitulate the known theory, using $N = n
\sim 10^{40}$ in (\ref{ee3}) we get
$$r = \sqrt{n}l \equiv L \sim 10^{-13}cm$$
$r$ now being a typical elementary particle, Compton length $L$.
Similarly, we can get from (\ref{ee5}), $M$ being the mass of an
elementary particle,
$$M = \frac{m}{\sqrt{n}} \sim 10^{-25}gm$$
It must be mentioned that in this theory, furthermore, $n \sim
\sqrt{N}$, where $N \sim 10^{80}$ is the number of elementary
particles in the universe (Cf.ref.\cite{bgsijmpa}).\\
It has also been shown that in the above approach there is a
pleasing correspondence with the usual Hawking-Beckenstein theory of
Black Hole Thermodynamics \cite{ijtpaugust}.\\
We can push the above consideration further. So far we have
considered only a coherent array. This is necessary for meaningful
physics and leads to the elementary particle masses and their other
parameters as seen above. Cercignani \cite{cer} had used Quantum
oscillations, though just before the dark energy era -- these were
the usual Zero Point oscillations, which had also been invoked by
the author in his model. Invoking gravitation, what he proved was,
in his own words, ''Because of the equivalence of mass and energy,
we can estimate that this (i.e. chaotic oscillations) will occur
when the former will be of the order of $G[(\hbar \omega )c^{-2}]^2
[\omega^{-1}c]^{-1} = G\hbar^2\omega^3 c^{-5}$, where $G$ is the
constant of gravitational attraction and we have used as distance
the wavelength. This must be less than the typical electromagnetic
energy $\hbar \omega$. Hence $\omega$ must be less than
$(G\hbar)^{-1/2}c^{5/2}$, which gives a gravitational cut off for
the frequency in the zero-point energy."\\
In other words he deduced that there has to be a maximum frequency
of oscillators given by
\begin{equation}
G\hbar \omega^2_{max} = c^5\label{e10}
\end{equation}
for the very existence of coherent oscillations (and so a coherent
universe). We would like to point out that if we use the above in
equation (\ref{e10}) we get the well known relation
\begin{equation}
Gm^2_P \approx \hbar c\label{B}
\end{equation}
which shows that at the Planck scale the gravitational and
electromagnetic strengths are of the same order. This is not
surprising because it was the very basis of Cercignani's derivation
-- if indeed the gravitational energy is greater than that given in
(\ref{B}), that is greater than the electromagnetic energy, then the
Zero Point oscillators, which we have called the Planck oscillators
would become chaotic and incoherent -- there would be no physics.\\
Let us now speak in terms of the background dark energy. We also use
the fact that there is a fundamental minimum spacetime interval,
namely at the Planck scale. Then we can argue that (\ref{B}) is the
necessary and sufficient condition for coherent Planck oscillators
to exist, in order that there be elementary particles which as noted
above has been shown to be the number of $n \sim 10^{40}$ coherent
Planck oscillators, and the rest of the requirements for the
meaningful physical universe. In other words gravitational energy
represented by the gravitation constant $G$ given in (\ref{B}) is a
measure of the energy from the background dark energy that allows a
physically meaningful universe -- in this sense it is not a separate
fundamental interaction.\\
It is interesting that (\ref{B}) also arises in Sakharov's treatment
of gravitation where it is a residual type of an energy
\cite{sakharov2,tduniv}.\\
To proceed if we use the expression for the elementary particle mass
$M$ seen above in terms of the Planck mass in (\ref{B}), we can
easily deduce
\begin{equation}
GM^2 \approx \frac{e^2}{n} = \frac{e^2}{\sqrt{N}}\label{C}
\end{equation}
where now $N \sim 10^{80}$, the number of particles in the
universe.\\
Equation (\ref{C}) has been known for a long time as an empirical
accident, without any fundamental explanation. Here we have deduced
it on the basis of the Planck oscillator model. Equation (\ref{C})
too brings out the relation between gravitation and the background
Zero Point Field or Quantum vacuum or dark energy. It shows that the
gravitational energy has the same origin as the electromagnetic
energy but is in a sense a smeared out effect over the $N$ particles
of the universe. In the context of the above considerations that
(\ref{C}) is deduced and not empirical as in the past, we can now
claim that (\ref{C}) gives the desired unified description of
electromagnetism and gravitation.\\
Finally the following may be pointed out: The work stemming from the
cosmology briefly referred to above, leads to a universal
acceleration, which is of the order of the pioneer anomaly, as has
been pointed out by the author. Indeed such a universal acceleration
is now being factored in as a reality \cite{smolin}.


\begin{thebibliography}{99}
\bibitem {narlikarcos} Narlikar, J.V. (1993). \emph{Introduction to Cosmology} (Cambridge University
Press, Cambridge), p.57.
\bibitem {tduniv} Sidharth, B.G. (2008). \emph{The Thermodynamic
Universe} (World Scientific), Singapore.
\bibitem {metz1} Metz, M, et al. (2009). \emph{The Astrophysical
Journal} 10.1088/0004-637X/697/1/269, 2009.
\bibitem {metz2} Metz, M. et al. (2009). \emph{Monthly Notices of
the Royal Astronomical Society} 10.1111/j.1365-2966.2009.14489.x,
2009.
\bibitem {physorg} http://physorg.com
\bibitem {petrosian} Petrosian, V. (2010). \emph{Astrophysical
Journal} February 10, 2010.
\bibitem {ag} Sawangwit, U. and Shanks, T. (2010). \emph{Astronomy and Geophysics} (Royal Astronomical Society)
Vol.51 (5), October 2010, pg.14ff.
\bibitem {mil1} Milgrom, M. (1983). \emph{APJ} 270, pp.371.
\bibitem {mil2} Milgrom, M. (1986). \emph{APJ} 302, pp.617.
\bibitem {mil3} Milgrom, M. (1989). \emph{Comm. Astrophys.} 13:4, pp.215.
\bibitem {mil4} Milgrom, M. (1994). \emph{Ann.Phys.} 229, pp.384.
\bibitem {mil5} Milgrom, M. (1997). \emph{Phys.Rev.E} 56, pp.1148.
\bibitem {ijmpa1998} Sidharth, B.G. (1998). \emph{Int.J.Mod.Phys.A}, 13 (15), 1998, p.2599ff.
\bibitem {cu} Sidharth, B.G. (2001). \emph{Chaotic Universe: From the Planck to the Hubble Scale}
(Nova Science, New York).
\bibitem {barrowparsons} Barrow, J.D. and Parsons, P. (1997). \emph{Phys.Rev.D.} Vol.55, No.4, 15
February 1997, pp.1906ff.
\bibitem {narfpl} Narlikar, J.V. (1983). \emph{Foundations of Physics} Vol.13, No.3, pp.311--323.
\bibitem {narburbridge} Narlikar, J.V. (1989). \emph{Did the Universe Originate in a Big Bang?}
in \emph{Cosmic Perspectives} Biswas, S.K., Mallik, D.C.V. and
Vishveshwara, C.V. (eds.) (Cambrdige University Press, Cambridge),
pp.109ff.
\bibitem {5} Sidharth, B.G. (2008). \emph{Ether, Space-Time and
Gravity} \textbf{Vol.3} Michael Duffy (ed.) (Apeiron Press, USA).
\bibitem {6} Sidharth, B.G. (2003). \emph{Chaos, Solitons and
Fractals} 18, (1), pp.197--201.
\bibitem {ruffinizang} Ruffini, R. and Zang, L.Z. (1983). \emph{Basic Concepts in Relativistic
Astrophysics} (World Scientific, Singapore), p.38.
\bibitem {hayakawa} Hayakawa, S. (1961). \emph{Suppl of Progressive Theoretical Physics Commemmorative Issue} 1961, pp.858-860.
\bibitem {nottalefractal} Nottale, L. (1993). \emph{Fractal Space-Time and Microphysics: Towards
a Theory of Scale Relativity} (World Scientific, Singapore), pp.312.
\bibitem {weinberggravcos} Weinberg, S. (1972). \emph{Gravitation and Cosmology}
(John Wiley \& Sons, New York), p.61ff.
\bibitem {uof} Sidharth, B.G. (2005). \emph{The Universe of Fluctuations} (Springer,
Netherlands).
\bibitem {uzan} Uzan, J.P. (2003). \emph{Rev.Mod.Phys.} 75, April 2003, pp.403--455.
\bibitem {bgsnc115b} Sidharth, B.G. (2000). \emph{Nuovo Cimento} \textbf{115B} (12), (2), pp.151ff.
\bibitem {bgsfpl} Sidharth, B.G. (2006). \emph{Foundations of
Physics Letters} 19 (6), 2006, pp.611-617.
\bibitem {sivaramfpl93} Sivaram, C. and Sabbata, V. de. (1993). \emph{Foundations of Physics Letters}
6, (6).
\bibitem {stave} Staveley-Smith, L. et al. (2002).
\emph{Astronomical Journal} Vol.24 (4), 2002, pp.1954-1974.
\bibitem {gold} Goldstein, H. (1966). \emph{Classical Mechanics} (Addison-Wesley, Reading,
Mass.), pp.76ff.
\bibitem {andersongrqc} Anderson, J.L. et al. \emph{xxx.lanl.gov/gr-qc/9808081}.
\bibitem {andersonphysrev} Anderson, J.D. et al. (2002). \emph{Phys.Rev.D} 65, pp.082004ff.
\bibitem {bgstest} Sidharth, B.G. (2006). \emph{Found.of Phys.Lett.} 19, (6), pp.611--617.
\bibitem {davis} Davies, P. (1989). \emph{The New Physics} Davies, P. (ed.) (Cambridge
University Press, Cambridge), pp.446ff.- New Physics
\bibitem {bgsgrav2} Sidharth, B.G., \emph{Ann.Fond.L.De Broglie} 30 (2), 2005, pp.151-156.
\bibitem {jacob} Jacob, M.,  (Ed.), ``Gauge Theory and Neutrino Physics'', North Holland, Amsterdam, 1978.
\bibitem {annales} Sidharth, B.G.,  Annales de la Fondation Lois de Broglie, Volume 27 No.2, 2002, 333-342.
\bibitem {bgsgrav} Sidharth, B.G.,  Nuovo Cimento, 116B (6), 2001, p.4ff.
\bibitem {ax} Sidharth, B.G., Nuovo Cimento, 117B, (6), 2002, pp.703ff.
\bibitem {fpl1}  Sidharth, B.G.,  Found.Phys.Lett., August 2002.
\bibitem {br2} Sidharth, B.G., Found.Phys.Lett., 15 (6), 2002, pp.577-583.
\bibitem {dirac4} Dirac, P.A.M.,  in ``Monopoles in Quantum Field Theory'', Eds. N.S. Craigie, P. Goddard and W. Nahm,
World Scientific, Singapore, 1982, p.iii.
\bibitem {nc} Sidharth, B.G.,  Nuovo Cimento, (6), 118B,  2002, 703ff.
\bibitem {ohanian} Ohanian, H.C., Rufkin, R., and Ruffini, R. (1994). \emph{Gravitation and Spacetime}
(W.W. Norton \& Company New York) 1994, pp.64ff.
\bibitem {berg} Bergmann, P.G. (1969). \emph{Introduction to the Theory of Relativity}
(Prentice-Hall, New Delhi), p248ff.
\bibitem {ar4} Sidharth, B.G.,  Gravitation and Cosmology, 4 (2) (14), 1998, p.158ff.
\bibitem {sakharov} Sidharth, B.G., Found.Phys.Lett., 16 (1), 2003, pp.91-97.
\bibitem {moriyasu}  Moriyasu, K.,  ``An Elementary Primer for Gauge Theory'', World Scientific, Singapore, 1983.
\bibitem {taylor} Taylor, J.C.,  ``Gauge Theories of Weak Interactions'', Cambridge University Press, Cambridge, 1978.
\bibitem {cr39}  Sidharth, B.G.,  Proceedings of the Fifth International Symposium on ``Frontiers of Fundamental Physics'', Universities Press,
Hyderabad, 2004 (In Press).
\bibitem {ijmpe} Sidharth, B.G. (2005). \emph{Int.J.Mod.Phys.E.} 14 (2), 2005, 215ff.
\bibitem {bgsijmpe} Sidharth, B.G. (2010). \emph{Int.J.Mod.Phys.E.}
19 (1), 2010, 79-91.
\bibitem {arxiv} Sidharth, B.G. \emph{arxiv 1008.2491}.
\bibitem {roe} Roe, B. \emph{http://physlinks.com}; to appear in
\emph{Phys.Rev.Letters}.
\bibitem {kiefer} Kiefer, C. (2004). \emph{Quantum Gravity}
(Clarendon Press, Oxford).
\bibitem {fpl2000} Sidharth, B.G. (2004). \emph{Found.Phys.Lett.}
{\bf 17}, (5), pp.503-506.
\bibitem {fpl152002} Sidharth, B.G. (2002). \emph{Found.Phys.Lett.}
15 (6), 2002, pp.577-583.
\bibitem {bgsijmpa} B.G. Sidharth. (2006). \emph{Int.J.Mod.Phys.A.} 21, (31),
December 2006, pp.6315.
\bibitem {ijtpaugust} Sidharth, B.G. (2009). \emph{Int.J.Th.Phys.} 48
(8), pp.2427-2431.
\bibitem {cer} Cercignani, C. (1998). \emph{Found.Phys.Lett.} Vol.11, No.2, pp.189-199.
\bibitem {sakharov2} Sakharov, A.D. (1968). \emph{Soviet Physics - Doklady} Vol.12, No.11, pp.1040--1041.
\bibitem {smolin} Smolin, L. (2006). \emph{The Trouble with Physics},
(Houghton Mifflin Company, New York).
\end{thebibliography}
\end{document}